# Advances in computational modeling approaches in pituitary gonadotropin signaling

Romain Yvinec[1], Pascale Crépieux[1], Eric Reiter[1], Anne Poupon[1] and Frédérique Clément[2,3]

[1] PRC, INRA, CNRS, IFCE, Université de Tours, 37380 Nouzilly, France.
[2] Inria, Université Paris-Saclay, 91120 Palaiseau, France.
[3] LMS, Ecole Polytechnique, CNRS, Université Paris-Saclay 91128 Palaiseau, France,

Corresponding author :
Romain Yvinec
romain.yvinec@inra.fr

**Abstract**

*Introduction* Pituitary gonadotropins play an essential and pivotal role in the control of human and animal reproduction within the hypothalamic-pituitary-gonadal (HPG) axis. The computational modeling of pituitary gonadotropin signaling encompasses phenomena of different natures such as the dynamic encoding of gonadotropin secretion, and the intracellular cascades triggered by gonadotropin binding to their cognate receptors, resulting in a variety of biological outcomes.

*Areas covered* We overview historical and ongoing issues in modeling and data analysis related to gonadotropin secretion in the field of both physiology and neuro-endocrinology. We mention the different mathematical formalisms involved, their interest and limits. We discuss open statistical questions in signal analysis associated with key endocrine issues. We also review recent advances in the modeling of the intracellular pathways activated by gonadotropins, which yields promising development for innovative approaches in drug discovery.

*Expert opinion* The greatest challenge to be tackled in computational modeling of pituitary gonadotropin signaling is the embedding of gonadotropin signaling within its natural multi-scale environment, from the single cell level, to the organic and whole HPG level. The development of modeling approaches of G protein-coupled receptor signaling, together with multicellular systems biology may lead to unexampled mechanistic understanding with critical expected fallouts in the therapeutic management of reproduction.

**Article highlights box:**

- Modeling of pituitary gonadotropin blood levels involves underlying endocrine feedback loops to shed light on the complex dynamical patterns of hormonal rhythms.
- Sophisticated statistical tools are needed to decipher the encoding of hormonal signals within the HPG axis.
- Dynamical modeling of the intracellular gonadotropin signaling networks leads to a mechanistic understanding of the gonadotropin action in a short time scale.
- Multi-scale modeling is needed to renew our understanding of the molecular, cellular, and physiological processes underlying the control of the reproductive function.
- Innovative approaches in drug discovery may arise from integrating the cellular and intracellular scales in a multi-scale modeling framework of the reproduction axis.

**1) Introduction**:

Systems Biology, which heavily relies on mathematical modeling, has long been recognized as an opportunity to discover new, more efficient and safer, drugs [1-3]. Systems Biology driven mathematical models allow one to understand the consequences of a local perturbation on the whole network behavior. For example, these models help understanding the effect of a drug, whose target is located in a precise cell type, on the physiological function this cell type participates in. These models are also very useful for determining the best targets within a physiological net [4]. As stated by the statistician George Box, "all models are wrong, but some are useful" [5]. Indeed, models have been proven to be useful in many different situations, from the network-based classification of metastatic cancers [6], or the susceptibility to metabolic disorders [7], to the study of anti-angiogenic therapies in cancer [8] and the mechanisms underlying neurodegenenerativon [9], just to name a few. However, modeling is not a straightforward task, and requires both the detailed knowledge of the studied system, and the selection of the most adapted mathematical formalism. In view of the complexity of the reproductive system, arising in particular from the multiple entangled levels of controls and its highly dynamic characteristics, it is understandable that the in-silico modeling approaches to drug discovery for reproductive biology are still in its infancy. Yet, a variety of mathematical tools have been used to help understanding the dynamics of the pituitary gonadotropin signaling and its perturbation, within the HPG axis. Here we present the state of the art in mathematical modeling applied to pituitary gonadotropin signaling, which could help designing better therapeutic solutions for reproductive disorders.

Pituitary gonadotropins are high molecular weight glycoproteins secreted by a specific type of pituitary cells, the gonadotrophs. In vertebrates, the pituitary gland is a pivotal organ within the neuroendocrine axes, linking the hypothalamus (and afferent connections) belonging to the central nervous system, to the peripheral target organs. In the hypothalamic-pituitary-gonadal (HPG) axis, a unique hypothalamic neuro-hormone, GnRH (gonadotropin-releasing hormone), exerts a direct and differential control onto the production and secretion of two pituitary hormones, FSH (follicle-stimulating hormone) and LH (luteinizing hormone) released by the same cell type.

FSH and LH control the double gonadal function of gametogenesis and steroidogenesis, through G protein-coupled receptors (GPCR) expressed specifically on somatic cells. In the testes, Leydig cells are endowed with LH receptors (LHCGR), while Sertoli cells express FSH receptors (FSHR) [10]. In the ovaries, granulosa cells of growing follicles bear FSHR, theca and granulosa cells from preovulatory follicles express LHCGR [11]. Gonadal steroid hormones like estradiol (E2), progesterone (P), and testosterone (Te) modulate in turn the secretion of pituitary LH and FSH, as well as hypothalamic GnRH, within entangled endocrine feedback loops. FSH secretion is further tuned by inhibin (a peptide hormone of gonadal origin) and, in addition, inhibin action is enhanced or toned down by local paracrine secretion of activin and follistatin, respectively [12]. In female, each reproductive cycle is characterized by a drop in FSH level, which is first suppressed by gonadal inhibin emanating from the whole cohort of terminally growing follicles, while the contribution of the dominant follicle(s) to E2 production further impact FSH secretion at the end of the follicular phase [11,13]. Note that there exist other gonadotropins that are not secreted from the pituitary. For instance, the human chorionic gonadotropin (hCG) has a placental origin, and interacts with the LHCGR of the ovary and promotes the maintenance of the corpus luteum during the beginning of pregnancy. We will deal almost exclusively with pituitary gonadotropins in this article.

The secretion patterns of GnRH, FSH and LH have remarkable dynamic features. GnRH is secreted in pulses, and its encoding as a pulsatile signal is a prerequisite to sustain gonadotropin secretion. As a result of the excitation-secretion coupling in gonadotroph cells, involving calcium-mediated exocytosis of secretion granules, LH is also secreted in a pulsatile manner. In contrast, FSH appears to be mainly secreted in a calcium-independent basal manner (Note that basal secretion does not imply time-constant FSH level nor constant secretion rate). In addition, in each species investigated so far, including non human primates [14,15], FSH and LH are secreted massively at the time of ovulation, under the control of the GnRH surge, occurring each ovarian cycle [16].

In the gonads, at the cellular scale, FSH and LH trigger, through their cognate GPCRs, multiple connected signaling pathways conveying hormonal signals. Multiple feedbacks and cross-talks contribute to a complex signaling network, which results in various cellular responses, spanning distinct spatial and time scales, from short-range membrane protein activation to sustained signaling and cell-cell communications [17].

Understanding the pituitary gonadotropin signaling is thus a challenging and multi-faceted issue, which does not only encompass the outcomes of ligand binding to the receptor, but also the dynamic encoding of the gonadotropin signal, subject to multiple feedback controls. In agreement to their role as pivotal endocrine players within the HPG axis, LH and FSH signaling networks are embedded in a multi-scale framework.

In the following, we will illustrate, with the help of selected instances, different approaches of modeling covering some of these facets and calling to various modeling formalisms.

**2) Encoding the pituitary gonadotropin signal**

We will first give an overview on different computational approaches aiming to reproduce and analyze the changes in the secretion rates of gonadotropins and the resulting, finely tuned, time-varying blood levels of pituitary gonadotropins in a physiological context.

These approaches deal with the encoding of gonadotropin levels as dynamic endocrine signals. These signals follow various dynamic regimes, mainly (quasi-)steady states or oscillatory regimes, and are characterized by a combination of properties (amplitude, frequency, duration) either considered on a rather long term (typically on the several-week term of an ovarian cycle, on a day-to-day basis), or on a shorter term (the several-hour term of secretion events considered as such).

In the former case, the main motivation is to be able to reproduce both qualitatively and quantitatively the patterns of FSH and LH levels, together with the levels of gonadal hormones. It amounts to setting in a proper mathematical music the multiple and entangled feedback loops at play within the HPG axis. In the latter case, the main motivation is to dissect the different steps (i.e. production, release, clearance) of the secretion events in order to get access to hidden endocrine information (typically the secretion rate in the cavernous sinus [18]), and to analyze possible differences in the secretion patterns specific to physiological conditions (i.e. age, gender, puberty) or pathological situations.

**2.1) Modeling the fluctuations in hormonal levels as the result of endocrine feedback loops within the HPG**

The observation of the periodic and coordinated fluctuations in hormone levels along the ovarian cycle, especially along the menstrual cycle in the human species, has been the initial driving force for the development of dynamic models of the interactions between the ovaries and pituitary gland. A long modeling history, continuing up to now [19], started from the core "push-and pull" concept associated with the FSH/estradiol feedback loop (FSH-stimulated estradiol secretion as opposed to estradiol-inhibited FSH secretion) studied as soon as in the early 1940s. The pioneering work of Lamport [20] was the first to investigate this question from a mathematical perspective and to introduce the natural mathematical formalism of ordinary differential equations (ODE) (Brief definitions and explanations of all technical mathematical terms are provided in a glossary in annex.) to tackle endocrine issues.

The two-dimensional linear ODE model analyzed by Lamport in 1940, modeling the estradiol level as a damped harmonic oscillator, failed to reproduce properly the pattern of estradiol oscillations. It was only twenty years later that this drawback was circumvented by Thompson [21], who separated the contribution of the growing dominant follicle (the future ovulatory follicle) to estradiol secretion in a three dimensional extension of the initial model. The model studied by Thompson is piecewise linear with reset "decision" functions (the size of the dominant follicle is reset to 0 after reaching a given threshold, representing the occurrence of ovulation), and its solution is an undamped oscillator.

Later developments, with a gold age in the early 1970s, have progressively complexified the model core structure, by adding nonlinearities in hormonal interactions, and other endocrine players, mainly progesterone and LH. Among others, a seminal instance is provided by the work of Bogumil *et al.* [22]. They considered a rudimentary pulsatile-like mode of LH secretion and distinguished the episodic secretion regime during the ovulatory surge from the constitutive FSH secretion or pulsatile LH secretion in the remaining of the ovarian cycle (the authors qualify as "tonic" and "phasic" the general versus surge secretion regime). The key ingredients of the gonadotropin dynamics mostly rely on three mechanisms: (1) FSH and LH secretions are controlled by nonlinear feedback terms mediating the effects of gonadal, and possibly adrenal, steroids; (2) FSH and LH removal rates in the plasma are linearly proportional to their plasma levels (first order process, or single exponential decay); (3) the LH surge occurs as a result of LH accumulation in the pituitary gland, controlled by stepwise functions that integrate through time the positive stimulus of steroids. Bogumil *et al* also introduced a more elaborate description of ovarian follicle dynamics, again with threshold-mediated transitions, including the FSH-induced recruitment of a cohort of growing follicles, as well as different development stages for the dominant follicle and corpus luteum. Altogether, combining the endocrine variables with the physiological ones, the model consists in 34 equations, involving a system of algebraic-integro-differential equations.

An alternative way to generate proper hormonal patterns while keeping a relatively tractable mathematical formalism is the use of delay differential equations (DDE). In the context of gonadotropin dynamic models, explicit constant time delays are often used in feedback terms entering the FSH and LH secretion rates. This is consistent with the effective latency in the feedback of gonadal hormones, upon pituitary gonadotropin hormone synthesis. For instance Clark *et al.* have designed in [23] an autonomous DDE system reproducing the average levels of FSH and LH, subject to the control exerted by estradiol, inhibin and progesterone, without introducing stepwise decision functions nor convolution integrals. Applying tools from Hopf bifurcation theory and performing numerical simulations, they have shown the existence of two distinct, locally asymptotically stable, periodic solutions. The first one is consistent with the hormonal patterns in normal menstrual cycles, and the other one to an abnormal cycle, that can correspond to a pathological endocrine status such as the Polycystic Ovarian Syndrome. Further developments and variants of this model have been proposed and are reviewed in [24]. The model can be refined by embedding additional variables to include more detailed biological knowledge. In particular, the distinction between two types of inhibin (inhibin A and inhibin B), which affect differentially the synthesis of FSH, and the role of Anti-Müllerian Hormone (AMH) on the developing follicles was included in [25]. These refinements allowed the authors to broaden the timescale of the model up to a lifelong model, and to investigate the impact of a putative AMH treatment on the onset of menopause.

The efforts in developing models of hormone interactions mainly regard the hypothalamic-pituitary-ovarian axis in women. Nevertheless, several approaches have tackled similar issues in different breeding [26] or laboratory species [27]. For instance, in line with the approach followed in [23], a model based on DDE has been designed in [26] for the bovine species, whose ovarian physiology is rather close to human ovarian physiology (roughly comparable duration of the ovarian cycle, existence of follicular waves and similar size of the ovarian follicles). Interesting approaches in fish species [28,29] have also been developed with an underlying motivation of comprehensive ecotoxicology.

The models discussed so far are physiologically-based pharmacokinetic (PBPK) models, where secretion/clearance mechanisms are included as building blocks, and hormone circulating levels are related to one another by means of linear or nonlinear functions. Such equations intend to describe in a simple (or even simplistic) way the recurrent growth and decline of the steroidogenic ovarian tissues, ovarian follicles and corpus luteum. This is possible with the help of logic, rather than dynamic smooth, functions to compensate for the inaccurate understanding of some processes and/or to skip too complex phenomena. The rationale behind the construction of such models is thus to embed the endocrine and physiological knowledge available at the time of the model design in a single mathematical framework. Beyond the question of gonadotropin dynamics, the concomitant study of such models has motivated important advances in the theoretical understanding of dynamical systems [30,31] in the mathematical physiology and mathematical biology communities.

As far as the hypothalamic-pituitary-testicular axis in men or males, a similar core structure as the E2-FSH push-and-pull concept arises from the GnRH-LH-testosterone feedback loop: GnRH-stimulated LH secretion, LH-stimulated testosterone secretion, and testosterone-inhibited GnRH secretion. This concept was first illustrated by Smith [32] who used the ODE formalism to derive a three-dimensional model analogous to the widespread feedback repression or Goodwin model (see [31]), and obtained an oscillatory solution that could describe periodic hormone patterns. We refer to [33] for recent developments based on a DDE formalism, and model outputs matching experimental observation under normal and perturbed conditions (such as castration and testosterone replacement).

**2.2) Computational approaches of the GnRH and LH pulse generator**

The endocrine system in male, as compared to females, has relatively simpler key features: a main gonadal player, testosterone, a static pool of steroidogenic cells in the gonad, no surge nor qualitative change in the secretion regime on the central side. Inhibin also affects FSH secretion in males [10], and oestradiol seems to contribute significantly to the steroid feedback exerted by testes onto the hypothamo-piuitary axis. However, the role of testosterone is prominent in the control of the GnRH-LH system, and has been the main focus of modeling approaches designed on the whole HPG scale. This has encouraged the design of more comprehensive modeling approaches facing explicitly the issue of GnRH and LH pulse modeling and its embedding within the whole HPG axis. The difficulty arises from the almost discontinuous character of pulses (with GnRH signal being encoded as a square wave) and the discrete nature of the times of pulse release events. To represent point events, one has to call to other mathematical formalisms than autonomous ODE or DDE systems: stochastic point processes [34], stochastic differential equations (SDE) [35], excitable dynamics in the framework of impulse ordinary differential equations [36], or stiff nonlinear ODE with several timescales [37].

Even if the variety of the involved formalisms hamper direct comparisons between these approaches, they all share the ability to generate time series of GnRH and/or LH inter-peak intervals (IPI) to follow on a short term basis (generally on the order of several hours) the pulse frequency and the frequency modulations related to physiological (e.g. circadian rhythmicity at puberty) or pathological (e.g. exposure to endocrine disruptors) conditions. In contrast, notwithstanding the specific mathematical formalism, they may differ according to the access to endocrine data.

Clinically-oriented studies can only make use of LH time series. Information on GnRH activity is not available, so that GnRH pulse times must be inferred from LH pulse times. In this framework, even if elaborate theoretical and computational works, mostly based on deconvolution methods, manage to reconstruct the most plausible (in a statistical sense) GnRH signal, no direct validation is possible. When such approaches are deployed with an objective of signal analysis such as pulse detection [38,39], it becomes very difficult to assess the validity of the detection results. Nevertheless such studies have a clear methodological interest since they have led to the development of useful statistical tools dedicated to challenging problems in data analysis. Also, the use of simulated secretion data, as performed in [40] and [41], can be of great help to assess the sensitivity to noisy and subsampled data, and the positive and negative predictive values of the detection method.

In contrast, in experimentally-oriented studies, one can take advantage of other sources of data retrieved in non-human primates, rodent or ruminant species. These are mainly multi-unit activity recordings (MUA, volleys of electrical activity recorded from the median eminence [42]), providing an electrophysiological correlate of the GnRH-induced LH pulses, and GnRH levels measured directly from the pituitary portal blood with a high time resolution.

MUA data have motivated the design of stochastic point process models to investigate the temporal structure of the HPG activity, which were used for instance in [34] to detect possible memory mechanisms in successive LH pulses, as well as a circadian rhythm.

The seminal work of Keenan *et al.* [35] has provided a SDE-based description of the male HPG incorporating a sophisticated point process for GnRH and LH pulses.
A Weibull renewal process, with an intensity depending on both GnRH and testosterone, represents the GnRH release times, and drives subsequent LH release times, subject to a deterministic refractory period and fixed time delay. The pulse shape follows a generalized gamma density. The pulse amplitude is derived from the detailed dynamics of the content of exocytosis secretion vesicles. These dynamics combine previous accumulation and new synthesis of LH or GnRH, ruled by stochastic rates (logistic functions integrating over time the testosterone and/or GnRH regulating activities, perturbed by an Ornstein-Uhlenbeck process). A continuous mode of synthesis for LH as well as for testosterone is also included, which eventually filters the upstream pulsatile signals. The comparison of the model outputs with data was later performed in [43] to investigate possible systematic changes affecting the dynamics of the GnRH-LH-Te axis in aging men. This SDE-based pulse generator has also been taken-up in [44] to represent pulse events in a female HPG model.

In the framework of pulse modulated systems, Churilov *et al.* have also incorporated the pulsatile secretion of GnRH into the ODE model initially proposed in [32], by introducing deterministic jump discontinuities as Dirac delta functions for the pulses in [45]. Both the firing times and amplitudes of the jumps are modulated by the testosterone level. The detailed mathematical analysis of the stability of periodic solutions is challenging; it is based on the design of an equivalent discrete time map (between any two GnRH pulses) and the study of its fixed points.

In the framework of excitable systems, Brown *et al.* have designed a mathematical neuroscience approach to represent in a compact and averaged way the dynamic neuron network underlying the GnRH/LH pulse generator [36]. They used the excitation property of a well-known model in electrophysiology, the FitzHugh-Nagumo (FHN) model. In this model, the input is a point stochastic process with varying amplitude and intensity, which generates the GnRH and LH pulses. The slow-fast structure of the FHN model was not fully exploited here. In contrast, timescale separation was at the source of the design and analysis of a multiple timescale model involving coupled FHN systems and coping not only with GnRH pulses, but also with the recurrent alternation between the pulse and surge regimes [37]. This GnRH pulse and surge generator is also able to capture the modulation of pulse frequency along an ovarian cycle and the effects of estradiol or progesterone bolus [46], in agreement with the whole corpus of biological knowledge drawn from GnRH portal blood data.

The objective of the deconvolution analysis is to recover the full secretion signal. Deconvolution-based methods are somehow close to the former PBPK-like models mentioned in the first part of this section, in the sense that they aim to explain the encoding of dynamic signals by the combination of a secretion mechanism with a clearance mechanism. The convolution ensues from the fact that, at a given time t, hormone molecules secreted at any time s less than t, and still not cleared-off by time t, can contribute to the current hormone level. We refer to [40] for an instance of non-parametric reconstruction of LH and FSH secretory rates, upon exogenous GnRH stimulation. In general, the results of the deconvolution procedure are dependent on the specific hypotheses underlying the secretory burst shape and clearance mechanisms, as well as on the statistical method and regularization scheme (see [47] for a review and comparison of different deconvolution-based methods).

Empirical detection methods have been provided outside the deconvolution framework, with the objective to provide one with a reliable sequence of IPIs, hence to detect the pulse peak times only, rather than to reconstruct the whole signal. By necessity, they often rely on ad-hoc threshold parameters and may generate false-positive and false-negative errors, mainly due to the fact that the effective, almost instantaneous, pulse times are almost never observed experimentally, so that the selection of the locally highest values as time peaks may be misleading. To circumscribe this pitfall, the algorithm proposed in [41] involves a mixture of local, semi-local and global criteria combined with basic PBPK notions (LH half-life). Its reliability has been deeply investigated on simulated and reference data, and it has been applied in different experimental contexts (see e.g. [48]). The field of pulse/peak detection is still very active from the methodological ground (see for instance the use of nonlinear diffusion equations reviewed in [49]).

The issue of the statistical estimation of signal features becomes even harder when one considers several linked data series, such as joint measurement of LH and FSH levels [50]. To decipher the inherent multi-hormone interactions gonadotropins are part of, Veldhuis *et al.* have combined peak detection algorithms, deconvolution-based methods and biomathematical modeling to reconstruct unobserved signals in a framework called ensemble models [51]. However, methods are still in active development, and no gold standard has been achieved yet. For instance, network inference and model-free approaches may also be used to unravel hormonal regulations [51,52].

*3)* **Decoding the pituitary gonadotropin signal at the cellular level**

Pituitary gonadotropins transmit their signal through GPCR receptors (FSHR and LHCGR [53]) in the gonadal cells, to control gametogenesis and steroidogenesis. The binding of gonadotropins to their cognate receptors triggers the activation of several intracellular signaling pathways and leads to global changes in gene transcription [54-56] and protein translation [57,58].

**3.1) Interaction network**

Signaling cascades activated by the gonadotropins mainly originate from the interaction of their receptors with Gαs and β-arrestin proteins. However, the large number of molecules participating in these reaction cascades, the numerous interconnections (with feedback) existing between these molecules, and the cross-talks between pathways, partly underlie the complexity of signaling networks [59].

A first step toward the understanding of the signaling dynamics is based on the notion of interaction network (see Figure 2), *i.e.* a graph summarizing the links (direct or indirect activation, inhibition, modulation, complexation, etc) between the molecules involved in the various signaling cascades. This graph may be derived from a careful analysis of biological experiments in various cell models.

Several attempts to characterize FSH-induced signaling networks have been made recently, in particular in Sertoli cells [17,60,61], in granulosa cells [17] or in cumulus cells [62]. See also [63] for a recent curation of the literature on FSH signaling. We are not aware of analogous results for LH-induced signaling networks, yet recent experimental works shed light on the different LH-dependent pathways in granulosa cells [64].

To face the complexity of interaction networks, and the always increasing volume of data, especially –omics data, computational tools are needed to gather and integrate information. For example, enrichment of -omics data following specific hormone stimulation is possible through the confrontation with large pathway databases, such as Ingenuity® Pathway Analysis or Cytoscape [65]. Such an approach has been used in [62] to identify key functions and pathways associated with a list of differentially expressed genes after FSH stimulation in bovine cultured cumulus cells. Logic-based inference may also complete possible reaction networks by abductive reasoning with perturbation experiments (knock-in, knock-out, siRNA etc) and has been applied to the FSH-induced signaling network in [66].

**3.2) Receptors Structures, trafficking and signaling bias**

Studies on the gonadotropin receptors biology may also inform on receptor trafficking, cross-talks between pathways, or identification of scaffolding or hub molecules [67]. Structural modeling helps to gain insights on specific mechanisms such as the activation of the receptor upon ligand binding, although the structures of the full-length gonadotropin receptors are currently not available. One instance of such an approach is provided by [68], where structure modeling helps to understand the different actions of the human luteinizing hormone (hLH) and the human chorionic gonadotropin (hCG) at their common receptor. Although hLH and hCG occupy the same binding site on the extracellular part of the receptor, 3D homology models, based on the known FSH/FSHR structure [69], predict that the subsequent interaction with the hinge region of the LHCGR receptor varies between the two hormones. See also [70] for a review on LHCGR and its structure-function relationships, [67] for a review on FSHR and [71] for recent advances in GnRH receptor structure. Structural modeling approaches may also reveal the architecture of molecular complexes involving the receptor [72], and predict direct interactions having important consequences within a signaling pathway.

Receptor oligomerization (formation of a protein complex that consists of a small number of receptors) also play an important role in the signaling networks. In particular, homodimers (complex made of two receptors of the same type) and heterodimers (complex made of two different receptors) have been shown for FSHR and LHCGR [73,74], which adds a new layer of complexity. Indeed, these different oligomers might induce different signaling, and may be selectively favored by a given ligand (natural or synthetic hormones, small molecules, etc). In particular, the signaling of these oligomers might be biased relative to each other, that is, the same set of signaling pathways is triggered, but with different relative efficacy. Signaling bias is now considered as a common feature to many GPCRs, and has profound therapeutic implications [75,76]. Recent evidence show that gonadotropin receptor signaling pathways can be biased by allosteric modulators or differentially activated in a context-dependent manner, leading to different cellular outcomes [77-80] like steroid productions. Functional selectivity also probably occurs at the GnRH receptor, which opens the way to interesting pharmacological opportunities [81].

From the modeling viewpoint, signaling bias has been revealed using classical equilibrium pharmacology models [82]. The so-called operational model is widely used to infer bias from dose-response data [83]. However, the current availability of fluorescence (or Förster) resonance energy transfer (FRET, mechanism describing energy transfer between two light-sensitive molecules, a donor and an acceptor) and bioluminescence resonance energy transfer (BRET, using bioluminescent luciferase as donor molecules instead of a light-sensitive molecule which has to be initially excited by illumination) techniques [84,85], as well as recent evidence of temporal signatures of bias signaling [86,87], motivate the use of dynamic modeling techniques to decipher the mechanisms underlying functional selectivity.

**3.3) Intracellular dynamic modeling**

Beyond the large size of the GPCR interaction networks and the numerous possibilities of cross-talks between signaling pathways, an additional and key layer of complexity comes from the dynamic properties of intracellular GPCR signaling. In fact, different types of stimulations (biochemical nature of the ligand, temporal pattern, dose etc) may activate the same molecules but at different subcellular locations or with distinct temporal signatures, leading to very different cellular responses [88,89].

Various modeling approaches have been used to represent GPCR signaling dynamics [90-93]. Once an interaction network has been built, a common framework based on ODE is used to dynamically represent the activation of signaling pathways. The interaction network is appropriately interpreted as a biochemical reaction network, and each reaction is translated into infinitesimal elimination/production rates for the concentration of its reactant/product, using the law of mass action.

A standard iterative workflow between model construction and experimental data is applied to address specific biological questions [94]. Once the dynamic model has been built, extensive optimization algorithms are used to estimate unknown parameters (kinetic rates, initial concentration, etc) [95], and dedicated statistical frameworks can be used for model selection [96].

To our knowledge, there are few detailed works on the intracellular modeling of pituitary gonadotropin intracellular signaling. Clément *et al.* [97] proposed a model of the dramatic increase in the efficiency of cAMP response along follicular development. A steady state analysis and parameter sensibility have been performed in the case of constant input, and numerical simulations were done for time-varying inputs. The dynamic regulation of p70S6 kinase, through both the cAMP and mTOR pathways, after either FSH or insulin stimulation, has been compared at two different developmental stages in primary rat Sertoli cells [98]. This model gave access to experimentally unavailable detailed quantitative description on p70S6 kinase complex phosphorylation mechanisms. At a coarser scale, Quignot and Bois [99] have used a dynamic model to simulate steroid synthesis under FSH stimulation and investigated the effects of endocrine disruptors on steroidogeneis, using both *in-vitro* and *in-vivo* experimental data on rat granulosa cells. Their model includes the CYP19 aromatase and Hsd17b1 enzyme, whose syntheses are known to be regulated by FSH through the cAMP signaling pathway. These enzymes control in turn the production of gonadal steroids.

One can expect that this research field will be rapidly growing, as much biological knowledge has been gathered recently (see sections above 2.1, and 2.2), and important and challenging open questions remain to be tackled from the modeling viewpoint. The temporal encoding of hormonal signals, as discussed in the first section of this review, is a powerful information-carrying mechanism, since cells may be able to behave as sophisticated decoding sensors [100].

Deciphering the impact of the pulsatile nature of the signal, and possible differential effects of the pulse frequency (as it is naturally the case for LH), on the downstream intracellular targets [101] is a particularly relevant issue in this context. The study of the respective contribution of the amplitude, duration and frequency of the signaling events will require the design of specific experimental setups, such as microfluidic devices for instance, allowing one to control accurately the encoding of the input signal, combined with the development of mathematical methods suited to the analysis of non-autonomous dynamical systems [102]. Up to now, the decoding of GnRH pulses by gonadotrophs has retained much more attention than that of LH pulses by gonadal cells. This is probably due to the fact that the pulsatility of GnRH is an absolute prerequisite for its biological action. Moreover, the frequency of GnRH pulses differentially controls the expression of the FSH and LH beta subunits and the release rates of FSH and LH (see review [103]). An additional interest in the framework of this review is that the decoding of GnRH pulses results in the encoding of gonadotropin signals. Hence the insight gained from models dedicated to GnRH pulse decoding will certainly be beneficial to the understanding of pituitary gonadotropin signaling as a whole. An impressive amount of modeling work has been dedicated to GnRH signaling (which falls out of the scope of this article, see a comprehensive review in [104]). These approaches raise a generic theoretical question: which network motifs are able to decode and discriminate different pulse frequency-coded signals and/or to preserve the frequency information downstream in the signaling cascades? Some steps forward have been made to answer this question through the careful analysis of low-dimensional ODE models corresponding to small network motifs [105,106]. Information theory and stochastic modeling have also been used to assess how reliable a signaling pathway can be in transmitting information from a given input into a given output [107]. Yet, much work is still needed to correctly embed these theoretical results in a realistic signaling network.

**4) Conclusion**

Many efforts in the study of gonadotropin signals have been put on modeling the fluctuations in circulating hormone levels on a day-to-day basis and on the statistical and mathematical analysis of the GnRH-driven LH pulse generator. This is mostly due to the wide availability of endocrine time series and their interest for clinical investigations. Mathematical modeling has proven to be useful and successful in bringing qualitative and quantitative information on hormonal rhythms. These approaches are still under active development, and models are being challenged to generate a variety of behaviors including relevant pathological situations. Current challenges in this direction include data fitting and statistical analysis of the measured signals (times series analysis), to supply information as accurate as possible on unobservable variables in natural and pathological conditions. Although the mathematical modeling approaches in reproductive pharmacology is still to be much more developed, we believe that these models will be particularly important for evaluating the consequences of treatments, either in pathological situations or for medically assisted procreation.

Yet, there will remain limitations in terms of mechanistic interpretation as long as the cellular and intracellular scales are not embedded in larger scale approaches. The GPCR community is very active currently, and has for instance developed new experimental tools that shed light on key mechanisms of GPCR trafficking. It is to be expected that the modeling of pituitary gonadotropin signals will benefit from a larger effort of the GPCR modeling community [108]. In turn, this will bring decisive tools for drug screening and development of innovative approaches in drug discovery for reproductive biology.

## 5) Expert opinion

In this review, we have illustrated some computational modeling approaches dealing with the proper assessment of FSH and LH release from rather blurred experimental data, the phenomenological "push-and-pull" like hormone dynamics ensuing from the endocrine dialogs between the gonads and pituitary, the detailed description of the molecular pathways triggered by FSHR and LHCGR, and some associated structural biology issues. Even if these issues have been dealt partially from the modeling viewpoint, there still remain many open questions.

To our opinion, the greatest challenge to be tackled is embedding gonadotropin signaling within its natural multi-scale environment, which encompasses the following different scales: the cellular level, the cell-to-cell level, and the cell population level.

At the cell level, there remain many challenges regarding the intracellular networks downstream the gonadotropin receptors. To date, most attention has been put on canonic second messenger pathway (such as cAMP), while it is now clearly established that there are several distinct signaling modules, such as $\beta$-arrestin-induced ones [76]. The systematic account of cross-talks between gonadotropin receptors, with growth-factor and/or steroid induced pathways, or even direct conformational effect of steroids, will also be decisive to yield more predictive computational models in drug discovery. An instance of functional interaction of great physiological impact is provided by the granulosa cells of terminally developing ovarian follicles in which FSHR and LHCGR coexist. FSHR and LHCGR signaling interact in different ways in granulosa cells. First the expression of LHCGR in granulosa cells is induced by FSHR signaling. Second, once granulosa cells are endowed with both FSHR and LHCGR, FSH and LH act in synergy on cAMP and steroid synthesis (see lower panel of Figure 3). Finally, in those cells there might be physical interactions between FSHR and LHCGR, yet such interactions remain to be assessed in vivo (there are evidence in *in vitro* devices [73,74]).

Systems biology approaches, able to aggregate biological knowledge in a single framework and predict effect of (physiological or pharmaceutical) alterations of these networks, are going to be key tools in drug discovery in reproduction. Concomitantly, methodological improvements will be needed as more involved formalisms are required. The spatialization of the signaling modules and actors (nucleus, cytoplasmic, scaffolding, endocytosis) has to be taken into account, as it can be associated with clearly different kinetics, hence different final biological outcomes [109]. For instance, a general mechanism yielding persistent cAMP signals triggered by internalized GPCR has been proposed and supported by a reaction-diffusion model [110]. Recently, this mechanism has been revealed for the LHCGR in mural granulosa cells by [111], and the authors have suggested that it could contribute to transmit signals up to the oocyte and be physiologically relevant for oocyte meiosis resumption. Spatial modeling approaches will then rapidly develop in the need of refined description of the signaling networks.

In addition, stochastic modeling approaches also becomes important. The recent study on the GnRH receptor at the single cell level [107] has allowed one to understand how negative feedback in the ERK signaling pathway provides an optimal information transfer at intermediate feedback levels. It is thus to be expected that advances in experimental tools will allow a more general study of signaling pathways at the single cell level, and that stochastic modeling will be helpful to decipher inherent biological variability and the intrinsic role of cell-cell response variations.

At the cell-to-cell level, the intercellular communication between cells expressing either LHCGR or FSHR is a key mechanism within the HPG axis.

A typical instance of coordination between FSH and LH signaling is the control of steroidogenesis in the somatic cells of ovarian follicles. LH-induced signaling in theca cells results in the production of androgens that are transferred to granulosa cells. In these latter cells, the androgens are converted into estrogens thanks to FSH-induced expression of the aromatase enzyme. This process is known as the two-cell two-gonadotropin model [112] (see upper panel of Figure 3 ). It has been considered in a phenomenological, coarse-grain manner in [44] [113], yet would deserve more dedicated studies, all the more since it can give rise to imbalanced steroidogenesis as encountered in pathological situations such as the Polycystic Ovarian Syndrome [114]. Again, spatial modeling may also help here to understand how the spatial distribution of FSHR and LHCGR within a follicle can affect the signaling processes [115].

A comparable coordination between FSH and LH signaling exists in the spermatogenesis process. When stimulated by LH, Leydig cells secrete testosterone, which together with FSH stimulate Sertoli cell activity and spermatogenesis [116]. However, up to our knowledge, there is no precise mathematical modeling to decipher this cell-cell communication. Undoubtedly, theoretical approaches will reveal nontrivial behaviors of this system, and will permit to shed light on possible alterations of its dynamics.

Finally, the ultimate challenge will be to embed gonadotropin-induced decision-making of individual cells into the mechanistic modeling of tissular or organic functions subject to systemic whole-body controls (cell population and tissue level). This means connecting fine-grain models designed at each level of the HPG. Some steps forward have already been made through the design and study of spatio-temporal multi-scale models for structured cell populations [117,118] in the context of ovarian follicle development. The gonadotropin-induced signaling is explicitly accounted for by control terms driving both the cell fates locally (proliferation, terminal differentiation, cell death) and the whole cell dynamics globally. The simulation and mathematical analysis of such models are really tricky and necessitate dedicated mathematical and computational developments. Even if the formulations of the control terms are based on biochemistry, they remain very compact. An even more challenging science front, both from the experimental and mathematical/computational grounds, consists in coupling the detailed dynamics of intracellular signaling networks induced by gonadotropins, with the dynamics of cell populations underlying physiological functions, which would open the way to unheard-of fallouts in systems biology and systems pharmacology.

**Acknowledgments.**

We are grateful to the anonymous reviewers for their helpful comments. We thank Danielle Monniaux for her careful reading of our manuscript and her invaluable inputs.

**Figure 1**: the hypothalamic-pituitary gonadal (HPG) axis

As any neuroendocrine axis, the reproductive (hypothalamic-pituitary gonadal) axis involves three anatomic levels : the hypothalamus and pituitary gland on the central side, the gonads (ovaries in females, testes in males) on the peripheral side. The gonadotropin-releasing hormone (GnRH) is secreted by endocrine neurons of the hypothalamus into a dedicated portal system, which preserves its encoding as a pulsatile signal up to its target cells within the pituitary gland, the gonadotrophs. In response to GnRH pulses, these cells are able to release both gonadotropins, the follicle-stimulating hormone (FSH) and the luteinizing hormone (LH). FSH and LH are released into the general blood flow and act upon the somatic cells of the gonads to sustain both gametogenesis (oocyte and sperm maturation) and steroidogenesis (production of steroid hormones such as progesterone, testosterone and estradiol). The main source of steroid hormones are Leydig cells in the testes, granulosa and theca cells in the ovarian follicles and luteal cells of the corpus luteum (the histological remnant of follicles after ovulation) in the ovaries. In turn, gonadal steroids affect the production and secretion of both the hypothalamic GnRH and gonadotropins, while FSH secretion is further modulated by gonadal inhibin. In addition, in females, at each ovarian cycle, the GnRH pattern switches from a pulsatile secretion regime to a surge regime resulting in a massive and prolonged increase in GnRH level, which triggers the LH ovulatory surge. The inserts represent the change in the estradiol and progesterone levels over a whole ovarian cycle, on a day-to-day basis (left side) and the changes in testosterone levels on an hour-to-hour basis (right side).

**Figure 2**: Schematic view of the FSHR signaling network

Representation of a model of the FSHR signaling network, gathering various signaling levels: trans-membrane receptors, transducer molecules, second messengers, effector molecules and kinases, transcription factors and translation modulators, mRNA and genes. Note that only a small subset of FSHR signaling network is shown. Roughly, four (linked) signaling pathways are represented: (i) Activation of calcium channels, mediated by either Gq or the adaptor protein, phosphotyrosine interacting with PH domain and leucine zipper 1 (APPL1); (ii) p38/ERK/PKA cAMP-dependent pathway; (iii) PI3K/AKT Gs-dependent pathway; (iv) mTOR/rpS6 , both Gs and β-arrestin dependent pathways. Note that these pathways are neither linear chain nor independent of each other, as multiple cross-talks and retroaction loops exist. LHCGR and EGFR are also represented at the cell membrane, to highlight possible cross-talks and receptor trans-activation (LHCGR is known to activate Gq, Gi, Gs, β-arrestin and Src pathways; EGFR activates PI3K and ERK). Finally, while some information on FSH-dependent gene transcription and protein translation is available in the literature, a current open question is the understanding of the comprehensive mechanistic link between the signaling pathways and gene expression level (thus, we have not represented any arrow here). Note also that the protein encoded by the gene AREG interacts with EGFR (thus adding another level of complexity in the cross-talks).

**Figure 3**: coordination between FSH and LH signaling

*Upper panel : the two-cell two-gonadotropin model*

In the ovaries, granulosa cells from ovarian follicles express FSH receptors while theca cells express LH receptors. The two cell types function in a coordinated manner, especially as far as the synthesis of steroid hormones is concerned. LH-induced signaling in theca cells results in the synthesis of testosterone from progesterone, catalyzed by the CYP17A1 enzyme. The aromatization of testosterone into estradiol by the CYP19A1 enzyme is in turn induced by FSH signaling in granulosa cells.

*Lower panel : FSH and LH synergic signaling in granulosa cells*

Although most of the time granulosa cells are deprived of LH receptors, they can become endowed with both gonadotropin receptors in specific physiological conditions, namely in follicles which have been selected for ovulation. In these follicles, LH receptor expression is induced by FSH signaling, which results in an enhanced cAMP output as well as a more efficient production of estradiol. Hence, compared to the other growing follicles, the dominant follicle gets the double advantage of becoming less dependent to FSH supply and contributing the most to the drop in FSH levels at the end of the follicular phase.

**List of abbreviations**

DDE: delay differential equations

E2: estradiol

FHN: FitzHugh-Nagumo

FSH: follicle-stimulating hormone

FSHR: FSH receptors

GnRH: gonadotropin-releasing hormone

GPCR: G protein-coupled receptor

hCG: human chorionic gonadotropin

hLH: human luteinizing hormone

HPG : hypothalamic-pituitary-gonadal

IPI: inter-peak intervals

LH: luteinizing hormone

LHCGR: LH receptors

MUA: multi-unit activity

ODE: ordinary differential equations

P: progesterone

PBPK: physiologically-based pharmacokinetic

SDE :stochastic differential equations

Te: testosterone

The figures are available at http://yvinec.perso.math.cnrs.fr/Publi/RYPCERAPFC_18_advances_gonado_figures.pdf

The supplemental mathematical notes are available at

    http://yvinec.perso.math.cnrs.fr/Publi/RYPCERAPFC_18_advances_gonado_supp.pdf